\documentclass[aps,pra,showpacs,twocolumn,superscriptaddress]{revtex4}

\usepackage{graphicx,color}
\usepackage{amsmath,amsthm,amsfonts,amssymb,bm}
\usepackage{times}
\usepackage{epsf}
\usepackage[colorlinks={true}]{hyperref}

\usepackage[T1]{fontenc}
\usepackage[utf8]{inputenc}
\usepackage{ulem}

\hypersetup{citecolor={blue}, filecolor={blue}, linkcolor={blue}, urlcolor={blue}}

\usepackage{graphicx}
\begin{document}

\title{Quantum coherence and uncertainty in the anisotropic XY chain}

\author{G. Karpat}
\affiliation{Faculdade de Ci\^encias, UNESP - Universidade Estadual Paulista, Bauru, SP, 17033-360, Brazil}
\author{B. \c{C}akmak}
\affiliation{Faculty of Engineering and Natural Sciences, Sabanci University, Tuzla, Istanbul, 34956, Turkey}
\author{F. F. Fanchini}
\email{fanchini@fc.unesp.br}
\affiliation{Faculdade de Ci\^encias, UNESP - Universidade Estadual Paulista, Bauru, SP, 17033-360, Brazil}

\begin{abstract}
We explore the local quantum coherence and the local quantum uncertainty, based on Wigner-Yanase skew information, in the ground state of the anisotropic spin-1/2 XY chain in transverse magnetic field. We show that the skew information, as a figure of merit, supplies the necessary information to reveal the occurrence of the second order phase transition and the completely factorized ground state in the XY model. Additionally, in the same context, we also discuss the usefulness of a simple experimentally friendly lower bound of local quantum coherence. Furthermore, we demonstrate how the connection between the appearance of non-analyticities in the local quantum uncertainty of the ground state and the quantum phase transitions does not hold in general, by providing explicit examples of the situation. Lastly, we discuss the ability of the local quantum coherence to accurately estimate the critical point of the phase transition, and investigate the robustness of the factorization phenomenon at low temperatures.
\end{abstract}


\maketitle

\section{Introduction}

In nature, there exist genuinely quantum transitions in the ground states of quantum many-body systems, resulting in qualitatively distinct phases of matter. Such phase transitions, which are purely driven by quantum fluctuations due to the Heisenberg uncertainty principle, are known as quantum phase transitions (QPT) \cite{qpt}. Although QPTs occur at absolute zero temperature as one of the parameters of the system is continuously changed across a critical point (CP) $\lambda_c$, they can also be observed at sufficiently low temperatures, where thermal fluctuations are not strong enough to excite the system from its ground state. QPTs are intrinsically connected with the energy level crossings taking place in the ground states of the quantum many-body systems, which typically lead to the appearance of non-analyticities in the ground state energy. In particular, while a discontinuity in the first derivative of the ground state energy is recognized as a first order QPT, a discontinuity or a divergence in the second derivative characterizes a second order QPT, in which case the first derivative of the ground state energy is continuous. On the other hand, there are also more involved types of QPTs \cite{contqpt}, which cannot be understood within this standard framework.

Quantum spin chains present several different kinds of quantum critical behavior, and thus serve as an natural playground for studying QPTs. In addition, when being subject to an external transverse magnetic field, they exhibit another fundamental aspect known as factorization \cite{factor}. This phenomenon is defined as the presence of a fully factorized ground state emerging at a particular value of the magnetic field, namely, at the factorization point (FP) $\lambda_f$. The occurrence of the factorization phenomenon has been demonstrated to be in connection with a change of symmetry in the ground state and also with a transition in the two-spin quantum correlations \cite{factorcon}.

Quantum systems possess correlations of genuine quantum nature, which are fundamental to numerous applications of quantum information science \cite{qit}. Since correlations among the constituents of many-body systems are closely related to the emergence of the QPTs and the factorized ground state, it is natural to investigate the link between these two phenomena and correlation measures. In fact, this relation has been recently studied from many different angles in quantum critical spin chains. Specifically, correlation measures such as entanglement \cite{entangle} and quantum discord \cite{discord} have been employed as figures of merit for the examination of the QPTs and factorization phenomenon \cite{qptdm,allt0x,allt0y,werlang,alltnot0,alltnot0steve,ssb,ssbandnott0}. Whereas most authors only considered the absolute zero temperature \cite{qptdm,allt0x,allt0y}, others examined the problem at finite temperatures as well \cite{werlang,alltnot0,alltnot0steve,ssb,ssbandnott0}.

The concept of skew information has been first introduced by Wigner and Yanase half a century ago \cite{WYSI}. The Wigner-Yanase skew information (WYSI) has several equally interesting interpretations in quantum physics discussed in the literature \cite{WYSIfisher,WYSIuncer,WYSIcorr,WYSIcoh}. On one hand, it can be adopted as a measure of the information embodied in a state that is skew to (not commuting with) an observable (a self-adjoint matrix) \cite{WYSI}. On the other hand, it can used as a measure of quantum uncertainty of an observable in a quantum state \cite{WYSIuncer}. Moreover, it has been very recently shown that WYSI constitutes a reliable measure of the coherence in a quantum state \cite{WYSIcoh}, where a simplified experimentally friendly alternative version of the coherence measure has been also introduced. Even though various types of bipartite correlations in the ground state of quantum spin chains have been studied largely in the context of QPTs, the relation between the coherence contained in single-spin or two-spin density matrices, and the QPTs and factorized ground state has not been discussed before.

In this work, we consider the anisotropic spin-1/2 XY chain in a transverse magnetic field due to the fact that this model exhibits both a QPT and a non-trivial factorized ground state. We first reveal how the QPT and factorization phenomenon are linked with the local quantum coherence (LQC) \cite{WYSIcoh}, as quantified by WYSI, in single-spin and two-spin reduced density matrices of the ground state of the spin chain. We examine the effects of simplification of the coherence measure on the information we can gain from it about the appearance of the QPT and factorized ground state. We show that the signal of the QPT manifests itself even in the experimentally accessible simplified version of the single-spin coherence, whose measurement does not require a full tomography of the state.  We also find out that although this simpler alternative still spotlights the CP of the QPT, the factorized ground state can no longer be detected in this setting. Moreover, by studying a novel quantum correlation measure, namely local quantum uncertainty (LQU) \cite{WYSIqc}, which is closely related to LQC, we discuss the consequences of the optimization involved in the evaluation of this measure for the identification of the CP and FP. Our results show that there exist non-analyticities appearing in LQU which in fact do not correspond to any critical behavior. Finally, we also take into account the effects of finite temperature to discuss how precisely can the coherence measure estimate the CP of the QPT, and the robustness of the factorized ground state against thermal effects.

This paper is organized as follows. In Section II, we introduce the anisotropic spin-1/2 XY chain in a transverse magnetic field, along with its analytic solution. In Section III, we study the single-spin and two-spin coherence based on WYSI in the ground state of the XY model. We discuss the relation of coherence to the QPT and factorization phenomenon both at absolute zero temperature and low temperatures. Section IV includes the summary of our results.

\section{Spin-1/2 XY chain in transverse field}

The Hamiltonian of the one-dimensional anisotropic spin-1/2 XY chain in a transverse magnetic field is given by
\begin{equation} \nonumber
 H=-\frac{\lambda}{2}\sum_{j=1}^{N} [(1+\gamma)\sigma^j_x \sigma^{j+1}_x +(1-\gamma)\sigma^{j}_{y}\sigma^{j+1}_{y}]-\sum_{j=1}^{N}\sigma^{j}_{z}
\end{equation}
where $\sigma^j_{x,y,z}$ are the usual Pauli operators at $j$th site, $\lambda$ denotes the strength of the inverse field, $\gamma \in [0,1]$ is the anisotropy parameter, and $N$ is the number of spins. While the Hamiltonian $H$ is in the Ising universality class for $\gamma\geq 0$ and corresponds to the Ising Hamiltonian in a transverse field when $\gamma=1$, it reduces to the XX chain for $\gamma =0$. This model has an order-disorder type second order QPT occurring at the CP $\lambda_c=1$, which separates a ferromagnetic and a paramagnetic phase. Furthermore, although the ground state of the XY model is in an entangled state in general, there exists a non-trivial factorization line corresponding to $\gamma^2+\lambda^{-2}=1$. Thus, the ground state becomes completely factorized at the FP,
\begin{equation} \label{factor}
\lambda_f=\frac{1}{\sqrt{1-\gamma^2}}.
\end{equation}

In the thermodynamic limit ($N\rightarrow\infty$), the XY model can be exactly diagonalized with the help of the usual technique of Jordan-Wigner and Bogoluibov transformations \cite{XYsol}. Due to the translational invariance of the system, the reduced density matrix of two spins at the sites $i$ and $j$ is dependent only on the distance between them, $r=|i-j|$. Considering that the XY Hamiltonian is also invariant under parity transformation (exhibits $Z_2$ symmetry), the reduced density matrix of two spins, having the distance $r$ between each other, is given by
\begin{equation} \label{twospindm}
 \rho_{0r} = \frac{1}{4}[I+\langle\sigma_z\rangle (\sigma^0_z+\sigma^r_z)]+\frac{1}{4}\sum_{\alpha=x,y,z}\langle\sigma^0_{\alpha}\sigma^r_{\alpha}\rangle
 \sigma^0_{\alpha}\sigma^r_{\alpha},
\end{equation}
where $I$ is the four-dimensional identity matrix. The magnetization and two-spin correlation functions are defined as \cite{XYsol}
\begin{equation} \nonumber
\langle\sigma^z\rangle =  -\int_0^{\pi} \frac{(1+\lambda\cos \phi)\tanh(\beta\omega_{\phi})}{2\pi\omega_{\phi}}d\phi,
\end{equation}
\begin{align} \nonumber
\langle\sigma_0^x\sigma_r^x\rangle = & \begin{vmatrix}
G_{-1} & G_{-2} & \cdots & G_{-r} \\
G_0 & G_{-1} & \cdots & G_{-r+1} \\
\vdots & \vdots & \ddots & \vdots \\
G_{r-2} & G_{r-3} & \cdots & G_{-1} \end{vmatrix}, \\ \nonumber
\langle\sigma_0^y\sigma_r^y\rangle = & \begin{vmatrix}
G_1 & G_0 & \cdots & G_{-r+2} \\
G_2 & G_1 & \cdots & G_{-r+3} \\
\vdots & \vdots & \ddots & \vdots \\
G_r & G_{r-1} & \cdots & G_1 \end{vmatrix}, \\ \nonumber
\langle\sigma_0^z\sigma_r^z\rangle = & \langle\sigma^z\rangle ^2-G_rG_{-r}.
\end{align}
where the function $G_r$ is given as follows:
\begin{align} \nonumber
 G_r= &\int_0^{\pi} \frac{\tanh(\beta\omega_{\phi})\cos(r\phi)(1+\lambda\cos \phi)}{2\pi\omega_{\phi}}d\phi \\ \nonumber
      &-\gamma\lambda\int_0^{\pi} \frac{\tanh(\beta\omega_{\phi})\sin(r\phi)\sin(\phi)}{2\pi\omega_{\phi}}d\phi,
\end{align}
and $\omega_{\phi}=\sqrt{(\gamma\lambda\sin\phi)^2+(1+\lambda\cos \phi)^2}/2$ with $\beta =1/kT$ being the inverse temperature. We should note that here we neglect the effects of spontaneous symmetry breaking (SSB) as it has been almost always done in the literature except for the few works that studied the impact of such effects \cite{ssb,ssbandnott0} in the ordered phase. Additionally, as most of the previous treatments, the ground state we deal with in this work is not the real ground state but rather the one that is widely known as the thermal ground state. Indeed, the thermal ground state corresponds to the limit $\beta\rightarrow\infty$ of the canonical ensemble,
\begin{equation}  \label{thermal}
\rho= \lim_{\beta\rightarrow\infty} \frac{e^{-\beta H}}{Z},
\end{equation}
where $Z$ is the partition function. We also remind that if the ground state is non-degenerate, then it is the same as the one obtained from Eq. (\ref{thermal}). However, in case of a degeneracy in the ground state, from Eq. (\ref{thermal}) we obtain an equal mixture of all possible ground states, which is what happens in the one-dimensional anisotropic XY model in transverse field.

\section{Coherence and local quantum uncertainty in the spin-1/2 XY chain}

The definition of the WYSI, which we adopt as a measure of coherence, is given by \cite{WYSI}
\begin{equation} \label{WYSI}
I(\rho, K)=-\frac{1}{2}\textmd{Tr}[\sqrt{\rho},K]^2,
\end{equation}
where the density matrix $\rho$ describes a quantum state, $K$ is an observable, and $[.,.]$ denotes the commutator. While the WYSI reduces to the variance $V(\rho,K)=\textmd{Tr}\rho K^2-(\textmd{Tr} \rho K)^2$ for pure states, it is upper bounded by the variance for mixed states. It is important to recognize that, unlike other indicators of uncertainty, WYSI remains unaffected from the classical mixing. Thus, it filters out the purely quantum uncertainty in a measurement. It has been very recently proven by Girolami that $I(\rho, K)$ given by Eq. (\ref{WYSI}) satisfies all the criteria for coherence monotones \cite{cohmon} and consequently can be used as a reliable measure of coherence \cite{WYSIcoh}. We note that the absence of coherence implies that no quantum uncertainty can be observed, and statistical errors are due to classical ignorance.

$K$-coherence of a quantum state is defined as the coherence carried by $\rho$ when measuring the observable $K$ (which is assumed to be bounded and non-degenerate) \cite{WYSIcoh}. Furthermore, in order to be able to rewrite the coherence measure $I(\rho, K)$ as a function of observables, Girolami has also introduced a simplified alternative version by dropping the square root from the density matrix $\rho$,
\begin{equation} \label{simpleWYSI}
I^L(\rho, K)=-\frac{1}{4}\textmd{Tr}[\rho,K]^2,
\end{equation}
which is a meaningful and an experimentally friendly lower bound, since it can be measured in an interferometric setup only by performing two programmable measurements, independently of the dimension of the quantum system. One can define the LQC for composite systems to quantify the coherence contained in them locally. For a bipartite system, the LQC is written as $I(\rho_{AB}, K_{A}\otimes I_{B})$ if we quantify the local coherence with respect to the first subsystem. Due to the fact that the systems we consider in our work is invariant upon exchanging two spins, the LQC remains also unchanged.

\begin{figure}[t]
\includegraphics[width=0.45\textwidth]{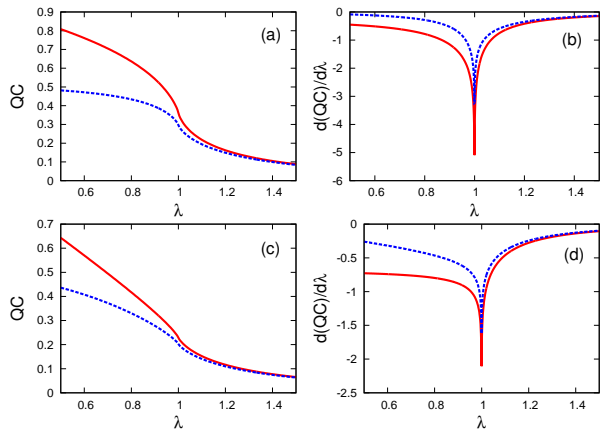}
\caption{Single-spin $\sigma_x$-coherence for $\gamma=0.5$ (a) and $\gamma=1$ (c), along with its first derivative (with respect to $\lambda$) for $\gamma=0.5$ (b) and $\gamma=1$ (d), as a function of $\lambda$. As the red solid line denotes the measure, the dashed blue line corresponds to its simplified version.}
\label{fig1}
\end{figure}

Another related concept is the LQU which is a full-fledged discord-like family of measures of purely quantum correlations \cite{WYSIqc}. In fact, LQU is nothing but an optimized version of the LQC over all possible local observables, that is,
\begin{equation} \label{measure}
U_A^\Gamma=\min_{K_A^\Gamma}I(\rho,K_A^\Gamma),
\end{equation}
where $\Gamma$ denotes the spectrum of $K_A^\Gamma$, and the minimization over a chosen spectrum of observables leads to a specific measure from the family. However, for a two qubit system, all the members of the family turn out to be equivalent. Then, the LQC can be analytically calculated as
\begin{equation} \nonumber
U_A(\rho_{AB})=1-\lambda_{\max}\{W_{AB}\},
\end{equation}
where $\lambda_{\max}$ is the maximum eigenvalue of the $3\times 3$ symmetric matrix $W_{AB}$ whose elements are given by
\begin{equation}  \nonumber
(W_{AB})_{ij}=\textmd{Tr}\left\{\sqrt{\rho_{AB}}(\sigma_{iA}\otimes I_B)\sqrt{\rho_{AB}}(\sigma_{jA}\otimes I_B)\right\},
\end{equation}
where indices $i,j=\{x,y,z\}$ are given for the usual Pauli operators. We note that Eq. (\ref{measure}) is normalized to one for maximally entangled pure states, and moreover, reduces to the linear entropy for any pure bipartite state.

\begin{figure}[b]
\includegraphics[width=0.45\textwidth]{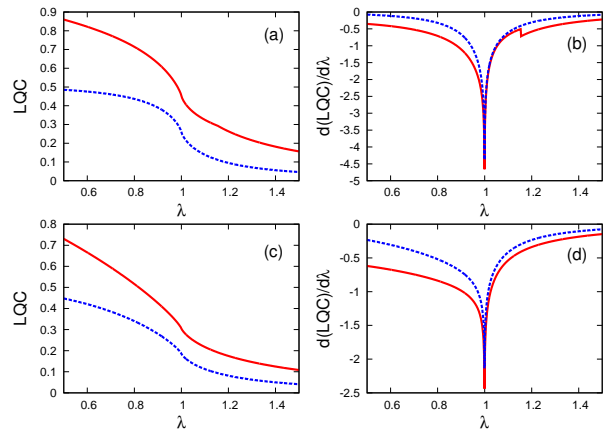}
\caption{Two-spin local $\sigma_x$-coherence for $\gamma=0.5$ (a) and $\gamma=1$ (c), along with its first derivative (with respect to $\lambda$) for $\gamma=0.5$ (b) and $\gamma=1$ (d), as a function of  $\lambda$. As the red solid line denotes the measure, the dashed blue line corresponds to its simplified version.}
\label{fig2}
\end{figure}

Having collected all the required tools for our analysis, we are now in a position to start our discussion regarding the relation between QPTs and factorization phenomenon, and quantum coherence based on WYSI. Let us first consider just a single spin from the whole chain. Since the XY model has translational invariance, all single-spin density matrices are the same and they are given by
\begin{align} \label{singlespindm}
\rho_0=\rho_i=\frac{1}{2}\begin{pmatrix} 1+ \langle\sigma^z\rangle & 0\\ 0 & 1- \langle\sigma^z\rangle \end{pmatrix},
\end{align}
where $\langle\sigma^z\rangle$ is the transverse magnetization, and the density matrix is written in the basis of the eigenvectors of $\sigma_z$. Note that, from this point on, we are working with the ground state in the limit $T\rightarrow0$ unless otherwise is stated.

In Fig. \ref{fig1}, we display the results of our analysis for the $\sigma_x$-coherence (coherence carried by $\rho_0$ when measuring $\sigma_x$) in the single-spin density matrix $\rho_0$ given by Eq. (\ref{singlespindm}) for two different values of the anisotropy parameter $\gamma$, namely for $\gamma=0.5$, and $\gamma=1$ which corresponds to the Ising model in transverse field. As can be observed from the plots of the derivatives of the measure shown in Fig. \ref{fig1}b and in Fig. \ref{fig1}d, while both the $\sigma_x$-coherence $I(\rho_0, \sigma_x)$ and its simplified alternative $I^L(\rho_0, \sigma_x)$ correctly spotlight both the location and the order of the CP of the second order QPT at $\lambda_c=1$ through a divergence in their first derivatives, no sign of the non-trivial FP can be seen for $\gamma=0.5$ at field $\lambda_f\sim 1.1547$. As, for $\gamma=1$, the FP would correspond to $\lambda_f\rightarrow \infty$ according to Eq. (\ref{factor}), we do not expect to see its signal in the plots. We should also remember that, we are analyzing the thermal ground state, thus the ground state is not pure despite being still separable at the factorization field $\lambda_f$. All the same, it is notable that even the simplified single-spin coherence measure given by Eq. (\ref{simpleWYSI}) detects the CP of the QPT since it can be determined without a full tomography of the state.

\begin{figure}[t]
\includegraphics[width=0.45\textwidth]{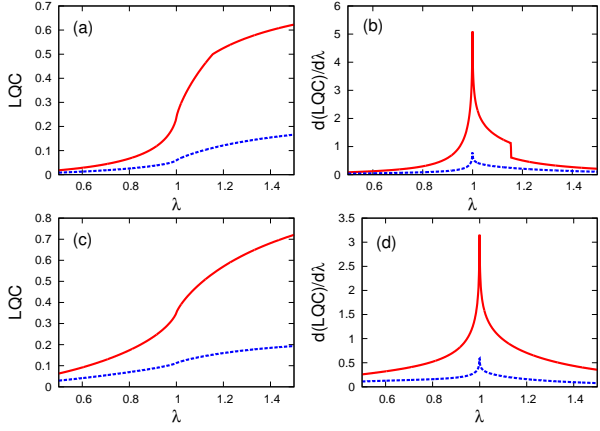}
\caption{Two-spin local $\sigma_z$-coherence for $\gamma=0.5$ (a) and $\gamma=1$ (c), along with its first derivative (with respect to $\lambda$) for $\gamma=0.5$ (b) and $\gamma=1$ (d), as a function of $\lambda$. As the red solid line denotes the measure, the dashed blue line corresponds to its simplified version.}
\label{fig3}
\end{figure}

The fact that there exists a relation between the appearance of a divergence in the derivative of the single-spin coherence of the ground state and the occurrence of the QPT can be understood within a general framework developed by Wu \textit{et al}. \cite{qptdm}. The energy of two spins at the sites $i$ and $j$ is given by
\begin{equation} \label{twositeenergy}
E(\rho_{ij})=\sum_{ij}\textmd{Tr}\left\{H_{ij}\rho_{ij}\right\},
\end{equation}
where $\rho_{ij}$ is the reduced density matrix of the spins and $H_{ij}$ is their reduced Hamiltonian whose summation over all sites restores the full Hamiltonian of the chain, $\sum_{ij} H_{ij}=H$. It is straightforward to obtain the first two derivatives of the two-site energy given by Eq. (\ref{twositeenergy}) with respect to the field $\lambda$ as
\begin{align} \nonumber
\frac{\partial E(\rho_{ij})}{\partial \lambda} = & \sum_{ij}\textmd{Tr} \left\{ \frac{\partial H_{ij}}{\partial \lambda}  \rho_{ij}\right\}, \\
\frac{\partial^2 E(\rho_{ij})}{\partial \lambda^2}= & \sum_{ij} \left[ \textmd{Tr} \left\{ \frac{\partial^2 H_{ij}}{\partial \lambda^2} \rho_{ij}\right\}+ \textmd{Tr} \left\{  \frac{\partial H_{ij}}{\partial \lambda}  \frac{\partial \rho_{ij}}{\partial \lambda}  \right\}  \right]. \nonumber
\end{align}
Considering that the derivatives of the reduced Hamiltonian are continuous with respect to the magnetic field $\lambda$, we realize that possible discontinuities in the derivatives of ground state energy have their roots at the elements of the reduced density matrices $\rho_{ij}$. Specifically, whereas a discontinuity in the first derivative of the ground state energy (a first order QPT) hints at a discontinuity in at least one of the elements of the reduced density matrix $\rho_{ij}$, a discontinuity or divergence in the second derivative of the ground state energy (a second order QPT) suggests a divergence of at least one of the elements of the derivative of the reduced density matrix $\partial \rho_{ij}/\partial \lambda$. Having this discussion in mind, it is rather straightforward to comprehend why two-spin or even single-spin coherence might be sufficient to pinpoint the CP of the QPT. However, it is very important to note that such a correspondence between the non-analyticities in physical quantities, that are functions of the reduced density matrix elements, and the CPs of QPTs does not always hold \cite{qptdm}. Depending on the mathematical properties of the considered quantity (correlation measures, coherence measures, etc.), it is possible that the CP of a QPT is not caught by a measure due to some unlucky coincidences. Conversely, we can also see non-analyticities in a measure which in fact do not correspond to any quantum critical behaviour. Therefore, whether such issues occur for the LQC and LQU is one of the questions that we will answer in this paper.

\begin{figure}[b]
\includegraphics[width=0.45\textwidth]{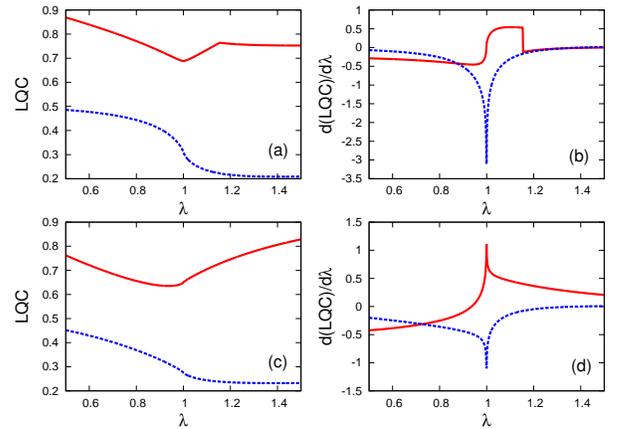}
\caption{Two-spin local $\sigma_y$-coherence for $\gamma=0.5$ (a) and $\gamma=1$ (c), along with its first derivative (with respect to $\lambda$) for $\gamma=0.5$ (b) and $\gamma=1$ (d), as a function of $\lambda$. As the red solid line denotes the measure, the dashed blue line corresponds to its simplified version.}
\label{fig4}
\end{figure}

We continue our investigation by exploring the two-spin LQC in the XY model, where we consider the nearest neighbor spins, i.e., $r=|i-j|=1$. Note that from this point on, we consider the local coherence meaning the observable acts only on one of the subsystems, that is, we evaluate $I(\rho_{AB}, K_{A}\otimes I_{B})$. Let us first examine the local $\sigma_x$-coherence contained in the reduced two-spin system $\rho_{01}$ given by Eq. (\ref{twospindm}). Fig. \ref{fig2} presents the outcomes of our analysis regarding the local $\sigma_x$-coherence in the ground state. It is evident that the results presented here seem very similar to those that are shown in Fig. \ref{fig1} for the single-spin $\sigma_x$-coherence in terms the link between the second order QPT at the CP $\lambda_c=1$ and the divergence in the derivative of the coherence. However, we notice that a new intriguing finite discontinuity shows up in the derivative in Fig. \ref{fig2}b at the field $\lambda\sim1.1547$, which is a result of the small kink appearing in Fig. \ref{fig2}a. This is nothing but the signal of the completely factorized ground state occurring at the FP $\lambda_f\sim1.1547$. It is worth to remark that it is rather unexpected to see a manifestation of the FP in the behavior of the coherence (even when ignoring the effects the SSB) since the WYSI has no direct relation to quantification of entanglement for mixed states. We also point out an important difference between the coherence measure based on the WYSI and its simplified version introduced by dropping the square root from the density matrix of the system. In particular, even though both the original definition and its simplified alternative might be equally useful in most regards, the simplified one, namely $I^L(\rho_0, \sigma_x)$, does not feel the existence of the factorized ground state at the FP.

\begin{figure}[t]
\includegraphics[width=0.45\textwidth]{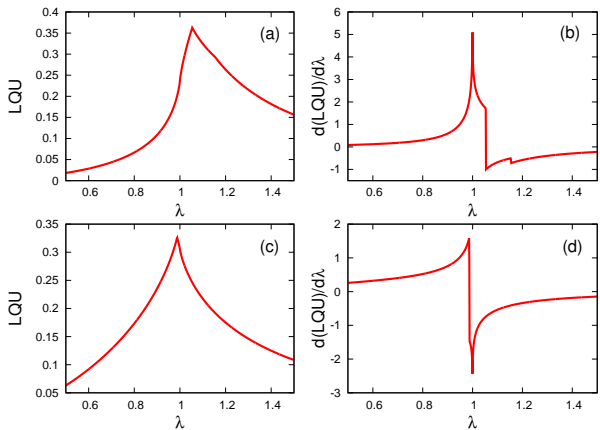}
\caption{Two-spin local quantum uncertainty for $\gamma=0.5$ (a) and $\gamma=1$ (c), along with its first derivative (with respect to $\lambda$) for $\gamma=0.5$ (b) and $\gamma=1$ (d), as a function of $\lambda$.}
\label{fig5}
\end{figure}

The reason behind this disagreement is without doubt the appearance of the square root in the definition of the WYSI. We stress that the emergence of the finite discontinuity in the derivative at $\lambda_f$ is not an accident, and can be seen for other values of the anisotropy parameter $\gamma$ as well. Having a closer look at the two-spin reduced density matrix, we realize that this discontinuity has its roots in the elements of $\sqrt{\rho_{01}}$, and is transferred from them to the LQC. Therefore, not only the WYSI but also the other physical quantities which are similarly based on $\sqrt{\rho_{01}}$, can pinpoint the FP $\lambda_f$. For instance, bipartite entanglement measures such as concurrence and entanglement of formation, which is itself a function of concurrence, have been studied in the ground state of the XY model. Interestingly, both of these measures also depend on $\sqrt{\rho_{01}}$ but, since they vanish at $\lambda_f$ due to the fact that even the thermal ground state is separable at the factorization field, the connection between the elements of $\sqrt{\rho_{01}}$ and the factorization phenomenon has not been explicitly realized. We emphasize that this correspondence is fundamentally different from what happens for the QPT since neither ground state energy nor any other thermodynamic quantity had a discontinuity at $\lambda_f$.

Next, we discuss the results of the same analysis for the local $\sigma_z$-coherence in the ground state. Note that the $\sigma_z$-coherence vanishes, as required, for the single-spin state $\rho_0$ as it is diagonal in the $\sigma_z$ basis. However, it is clear that this is no longer true for the LQC. Fig. \ref{fig3} displays the local coherence carried by the nearest neighbor two-spin density matrix $\rho_{01}$, when measuring the observable $\sigma_z$, and also its derivative. We observe that the LQC in this case, despite behaving quantitatively differently from the $\sigma_x$-coherence for both the XY model ($\gamma=0.5$) and the Ising model ($\gamma=1$), leads us to the same conclusion about the CP of the QPT and the FP.

We finish our examination of the LQC with the local $\sigma_y$-coherence in the ground state of the XY chain. Comparing Fig. \ref{fig4}a to what we observe in \ref{fig3}a and \ref{fig2}a, we see an unexpected behavior, that is, the coherence in this case has a minimum at the CP $\lambda_c=1$, which is also reflected to the derivative of the measure shown in \ref{fig4}b. As a consequence, the second order QPT cannot be detected as a divergence in the first derivative of the LQC. This is actually the result of an unlucky coincidence, which apparently cancels out the divergence in the derivative at the CP, occurring only for this particular observable and in case of $\gamma=0.5$. In fact, checking \ref{fig4}c and \ref{fig4}d, it is clear that the coherence exhibits the expected behavior for $\gamma=1$. On the other hand, the simplified $\sigma_y$-coherence does not suffer from this issue at any value of the anisotropy parameters $\gamma$. Thus, the example we presented here is not a systematic issue related the coherence measure based on WYSI for identifying the CP of the QPT. We also point out that the FP at $\lambda_f\sim 1.1547$ manifests its presence in the coherence measure again through a discontinuity in the first derivative.

\begin{figure}[b]
\includegraphics[width=0.48\textwidth]{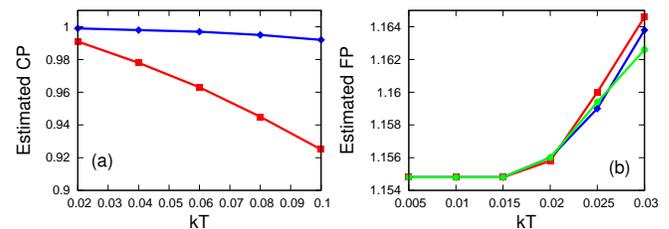}
\caption{(a) The Critical point estimated by single-spin $\sigma_x$-coherence (red line) and its simplified version (blue line) as a function of the temperature for $\gamma=0.5$. (b) The factorization field estimated by local two-spin $\sigma_x$-coherence (red line), $\sigma_y$-coherence (blue line) and $\sigma_z$-coherence (green line) as a function of time for $\gamma=0.5$.}
\label{fig6}
\end{figure}

Having discussed the LQC in the ground state of the XY chain case by case, we now turn attention to what the LQU, which is in fact the optimized version of the LQC over the set of all possible observables, has to say about the QPT and the factorization phenomenon. Fig. \ref{fig5} displays the behavior of the LQU and its derivatives for the cases of $\gamma=0.5$ and $\gamma=1$. Apart from the appearance of the divergence at the CP $\lambda_c=1$ and the finite discontinuity at the FP $\lambda_f\sim 1.1547$ in the first derivative of the measure, we also observe two new pronounced maxima in Fig. \ref{fig5}a and in Fig. \ref{fig5}c, corresponding to finite discontinuities in the derivatives shown in \ref{fig5}b and \ref{fig5}d. Indeed, the XY model has neither a QPT nor a factorized ground state at these points. A closer glance at the measure reveals the reason behind this: due to the optimization procedure in the definition of the LQU, there might occur sudden changes of the optimal observable, as we vary the magnetic field continuously. Particularly, in both plots Fig. \ref{fig5}a and Fig. \ref{fig5}c, the optimal observable jumps from $\sigma_z$ to $\sigma_x$ at these two new maxima. Hence, it is important to mention that the non-analyticities in the derivative of the LQU here do not come from the elements of the two-spin reduced density matrix $\rho_{01}$ but rather stem from the definition of the LQU naturally, and thus, should not be related to a quantum critical behavior.

Lastly, we briefly explore the ability of the LQC to correctly estimate the CP of the QPT at finite but sufficiently low temperatures, which might be considered effectively zero since the thermal fluctuations in this case are not strong enough to excite the system from its ground state. In spite of the fact that singular behaviour of the LQC disappears as the temperature rises, we might still estimate $\lambda_c$ to a reasonable accuracy. Additionally, we also perform a similar analysis for the FP to check the robustness of the emergence of factorization phenomena at finite temperatures. Our strategy can be summarized as follows: since, at finite temperature, a divergence in the first derivative of the LQC at $T = 0$ will be replaced by a local maximum or minimum about the singular point, we search for this extremum to estimate the CP. On the other hand, if the first derivative is discontinuous, then we look for an extremum in the second derivative of the LQC \cite{werlang}.

In Fig. \ref{fig6}a, we show the performances of the single-spin $\sigma_x$-coherence (red line) and its simplified version (blue line) in estimating the CP of the QPT. It is important to note that experimentally friendly alternative is a very accurate estimator of the CP of the QPT for $\gamma=0.5$ even at relatively high temperatures. Moreover, Fig. \ref{fig6}b demonstrates the outcome of the same analysis for the FP considering the $\sigma_x$-coherence (red line), $\sigma_y$-coherence (blue line) and $\sigma_z$-coherence (green line). We emphasize that the factorization phenomena is robust against the thermal effects until a certain temperature is reached. In fact, quantum discord has been also studied to investigate the same problem \cite{ssbandnott0}. However, the detection of the FP requires the evaluation of discord in the two-spin reduced system for more than one value of $r=|i-j|$. In particular, discord signals the FT through the intersection of lines plotted for different spin distances, i.e., it has the same value independent of the distance between the spins. Thus, the fact that the LQC serves the same purpose only considering the nearest neighbors, might be considered as an advantage over quantum discord. Note that, in case of finite XY chain, the robustness of the factorization phenomena can be explained in terms of the difference between the excited energy levels \cite{alltnot0steve}.

\section{Conclusion}

In summary, we have presented a systematic analysis of the relation of the QPT and factorization phenomenon, taking place in ground state of the anisotropic spin-1/2 XY chain in transverse magnetic field, to the LQC and LQU contained in the single-spin and two-spin reduced density matrices of the thermal ground state. On one hand, we show that an experimentally accessible simple measure of coherence based on WYSI can identify the CP of the second order QPT in the XY model, even when only a single-spin reduced system of the chain is considered. Moreover, the single-spin coherence remains as a very accurate estimator of the CP even at relatively high temperatures. On the other hand, our results clearly demonstrate that the connection between the QPTs and non-analyticities occurring in the LQC and LQU should not be taken for granted in general. For instance, the optimization procedure in the definition of the LQU might give rise to singularities in the behavior of the measure, due to a sudden change of the optimal observable, which do not correspond to any quantum critical behavior. Indeed, the examples we presented here should be considered as a particular case of similar situations that might be observed for all physical quantities involving an optimization procedure in their definitions \cite{werlang}.

Furthermore, we have shown that despite the fact that the LQC and LQU have no direct relation to any measure of entanglement for mixed states, they both show the signal of the completely factorized ground state in the XY model, due to the fact that their definitions are based on the WYSI. By further investigating this correspondence, we have demonstrated that the finite discontinuities emerging in the derivatives of LQC and LQU at the FP are actually transformed to the measures from the elements of the square root of the two-spin density matrix. This fact also explains why the simplified coherence measure based on $\rho$ instead of $\sqrt{\rho}$ does not tell anything about the factorization phenomenon. Lastly, we have examined the robustness of the factorization phenomenon in terms of the LQC at finite temperatures, and demonstrated that, as long as we consider sufficiently low temperatures, the LQC can still identify the factorized ground state.

\begin{acknowledgments}
The authors would like to thank Steve Campbell for his valuable comments on the manuscript. GK is supported by S\~{a}o Paulo Research Foundation (FAPESP) under grant number 2012/18558-5 and FFF under grant number 2012/50464-0. FFF is also supported by the National Counsel of Technological and Scientific Development (CNPq) under grant number 474592/2013-8 and by the the National Institute for Science and Technology of Quantum Information (INCT-IQ) under process number 2008/57856-6. B\c{C} is supported by the Scientific and Technological Research Council of Turkey (TUBITAK) under Grant No. 111T232.
\end{acknowledgments}

\end{document}